\documentclass[pre,twocolumn,showpacs,floatfix]{revtex4}
\usepackage{graphicx,amsmath,amssymb,url}
\usepackage{color}
\usepackage{dcolumn}

\begin{document}
\title{Local interaction scale controls the existence of a non-trivial optimal critical mass in opinion spreading}
\author{Zhi-Xi Wu\email{zhi-xi.wu@physics.umu.se}}
\affiliation{Department of
Physics, Ume{\aa} University, 901 87 Ume{\aa}, Sweden}
\author{Petter Holme}
\affiliation{Department of
Physics, Ume{\aa} University, 901 87 Ume{\aa}, Sweden}
\affiliation{Department of Energy Science, Sungkyunkwan University, Suwon 440--746, Korea}

\begin{abstract}
We study a model of opinion formation where the collective decision of group is said to happen if the fraction of agents having the most common opinion exceeds a threshold value, a \textit{critical mass}. We find that there exists a unique, non-trivial critical mass giving the most efficient convergence to consensus. In addition, we observe that for small critical masses, the characteristic time scale for the relaxation to consensus splits into two. The shorter time scale corresponds to a direct relaxation and the longer can be explained by the existence of intermediate, metastable states similar to those found in [P.\ Chen and S.\ Redner, Phys.\ Rev.\ E \textbf{71}, 036101 (2005)]. This longer time-scale is dependent on the precise condition for consensus---with a modification of the condition it can go away.
\end{abstract}
 \maketitle

\section{Introduction}\label{intro}

The study how human behavior form large-scale phenomena in society, from the spread of languages to election results, has a rather long and very interdisciplinary history. Several works in this tradition take statistical physics as their starting point~\cite{Haken1975rmp,Haken1983book,Stauffer2006book} (see the Ref.~\cite{Castellano2009rmp} for a recent review). Perhaps the most well-studied social spreading phenomena, apart from disease epidemiology, is opinion formation~\cite{Sznajd2000ijpmc,Galam1999pa,Galam2002epjb,
Krapivsky2003prl,Chen2005pre,Li2006pre,marsili,Lambiotte2007pre,grnld,
Klimek2008epl,Holme2006pre,Guan2007pre,Yang2009pre,Castellano2009pre}.

A popular model of opinion formation is majority rule model, in which agents update their opinions based on a local majority rule~\cite{Galam1999pa,Galam2002epjb,Krapivsky2003prl}. This rule is based on a common decision method for public debates. In fact, the majority rule serves as an efficient and fair way to solve problems of conflict and competition in many realistic social situations, and constitutionalized in many democracies. One concrete example is the presidential elections in the United States: in most of the states, the election is implemented by using a ``winner-takes-it-all'' rule, i.e., the candidate receiving the most votes in that state gets all the electors. Another example is public referenda---as a result of the Irish Lisbon II referendum, to take a recent example, 67\% voting in favor led Ireland to adopt the Lisbon Treaty (concerning, among other things, to legislate a majority voting system for some decisions in the European Union). These examples illustrate the common practice to settle a decision when a fraction of a population exceeds a certain threshold. Threshold mechanisms as outlined above do not only appear in human populations but also in economy and biology~\cite{schelling,Watts2002pnas}.

In a recent paper, Szolnoki and Perc studied the impact of critical mass on the evolution of cooperation in spatial public goods games~\cite{Szolnoki2010pre}. They found that the level of cooperation is highest at an intermediate value of the critical mass (the minimal number of cooperators required to harvest benefits), which is robust to variations of the group size and the interaction network topology. Inspired by~\cite{Szolnoki2010pre}, we here investigate how the size of critical mass within a group affects the opinion formation in the context of majority rule model. Most previous studies of majority rule model with binary opinions, either defined on completely connected graph (mean-field limit)~\cite{Galam1999pa,Galam2002epjb,Krapivsky2003prl}, $d$-dimensional regular lattice~\cite{Chen2005pre}, small-world networks~\cite{Li2006pre}, or on networks with strong degree heterogeneities or with community structure~\cite{Lambiotte2007pre}, assume simply that all the agents in a group will adopt the majority opinion in a sense that the critical mass can be regarded as one half. In some real life situations, however, a quorum far larger than one half is necessary to pass a resolution. Since consensus is the only possible final state in the majority rule model~\cite{Krapivsky2003prl,Chen2005pre}, we will focus our attention on whether the magnitude of the threshold value impacts the efficiency of the system to reach consensus, and if so, how it does that.

\section{Model}\label{model}

\begin{figure}
\includegraphics[width=\linewidth]{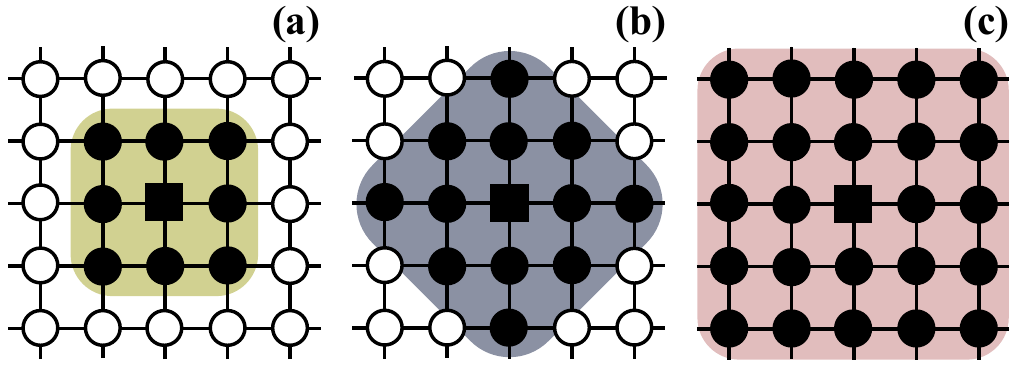}
\caption{(Color) Illustration of three different sizes of local-influence groups on a square lattice: the agent located on the central site can join a local-influence group of size (a) $9$, (b) $13$, and (c) size $25$.}\label{lattice}
\end{figure}

In our threshold majority rule model, a population of $N=200\times200$ agents are situated on a square lattice with periodic boundary conditions. Initially, each agent holds one of the two opinions $+1$ or $-1$. At each iteration, an agent is selected at random and with it a surrounding local-influence group of $G$ agents. The focal agent and its surrounding is then updated. Fig.~\ref{lattice} illustrates the three different local-influence groups that we explored in the present study. The interactions within the group can lead to a consensus between the agents---if the majority opinion exceeds a critical value $n_0\in(G/2,G)$, we let all agents in the local-influence group take the majority opinion. The lower bound of the interval ensures that the majority rule is fulfilled---a decision is taken only if a majority is in favor of it. The upper bound guarantees that the system will converge eventually. If a consensus is not reached, the opinions are randomly redistributed within the local-influence group. For convenience, we refer these two operations as majority-rule process and opinion-reshuffle process, respectively. The above procedure is repeated until a final consensus is accomplished. When $n_0=G/2$ our model reduces to the common majority rule model. Note that our model is also different to~\cite{Klimek2008epl}, wherein an agent is convinced if there is at least a fraction $q$ of its neighbors sharing the same opinion. In contrast to single agent updating in~\cite{Klimek2008epl}, we adopt group updating which makes consensus more accessible.

\begin{figure}
\includegraphics[width=0.9\linewidth]{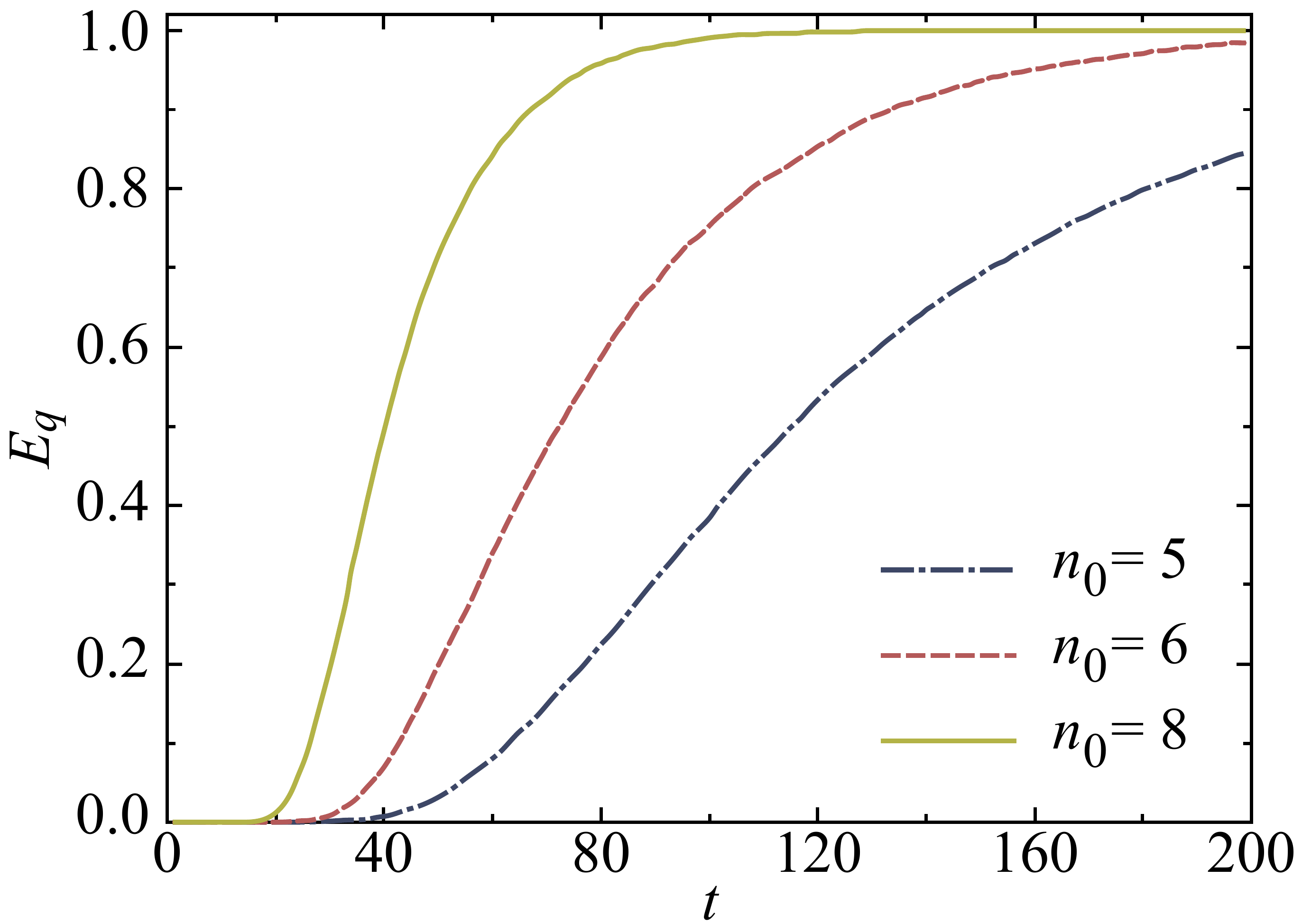}
\caption{(Color) Average exit probability $E_q$ as a function of the relaxation time $t$ (in units of Monte Carlo step per site) for local-influence group of size $G=9$ and different critical mass $n_0=5$, $6$, $7$, $8$. Initially, a fraction $q=52\%$ agents are assigned the opinion ``+1''. Each Monte Carlo time step corresponds to a sweep of group opinion updating in terms of a random sequence. The data are averages from 5000 independent realizations. }\label{fig1}
\end{figure}

\section{Results and Analysis}\label{results}

In numerical simulations, we fix the interaction group size $G$, and vary the value of $n_0$. Initially, the states are set to contain a randomly distributed fraction $q$ of ``+1'' and $1-q$ of ``-1'' opinions. The key quantity we measured is the average exit probability $E_q$ (that the system ends with homogeneous opinion state starting with a fraction $q$ of ``+1'') as a function of the relaxation time $t$, whose magnitude weighs the capability of the system to achieve consensus. For $q=0.5$, we find that in some realizations, the system is trapped in long-lived transient states, which evolve very slowly. To suppress such phenomenon, for each system $(G,n_0)$, we start from a state slightly favoring one of the opinions, say ``+1'', and iterate the system step by step. Under such setting, the initial minority can in principle dominate the whole system. We have checked that all the qualitative properties of the results presented below do not change if starting with equal fraction of the two binary opinions.

\begin{figure}
\includegraphics[width=0.9\linewidth]{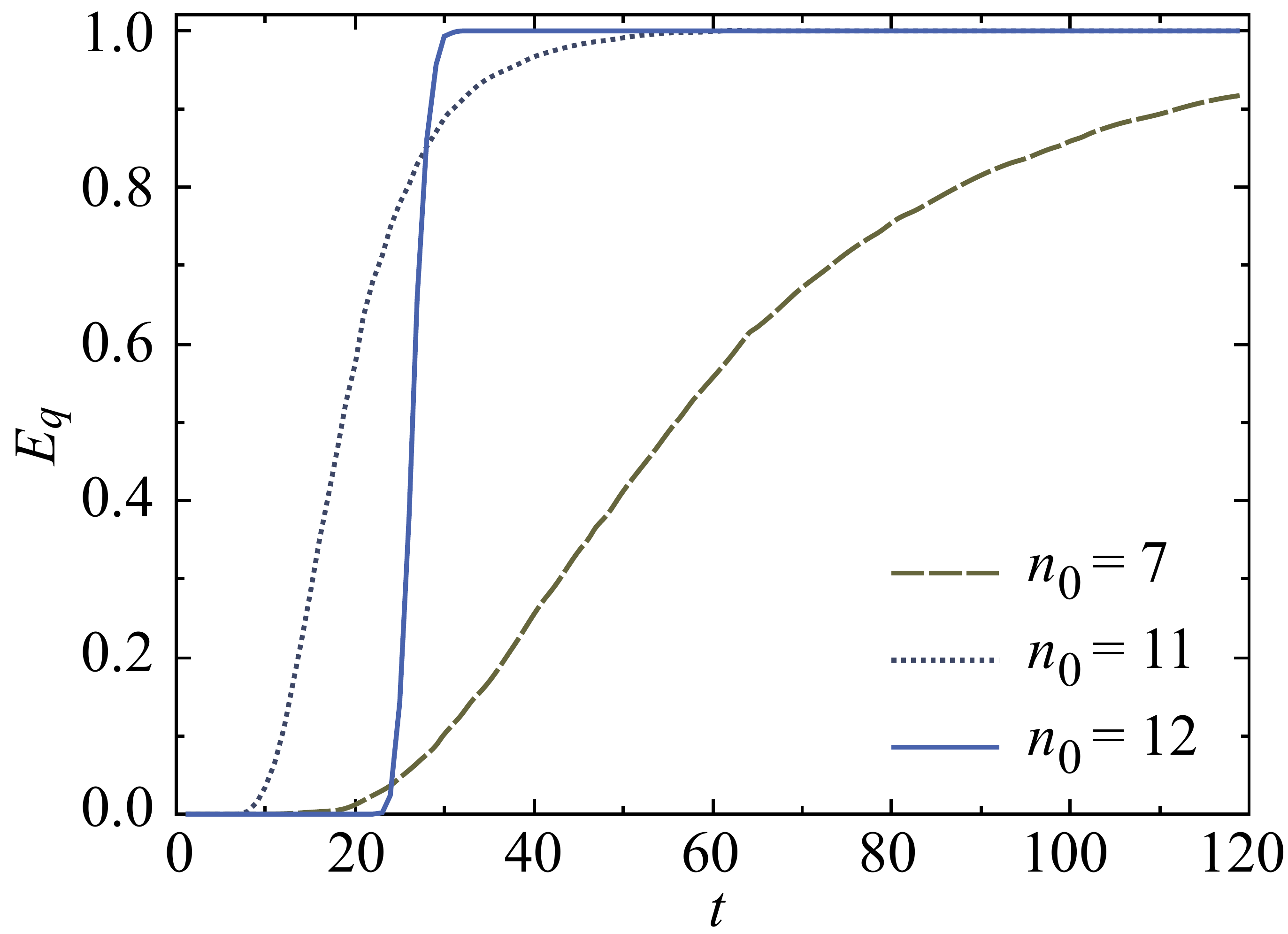}
\caption{(Color) Average exit probability as a function of the relaxation time, similar to Fig.~\ref{fig1}, but for a local-influence group of size $G=13$ and critical masses $n_0=7$, $11$, $12$. Note that $E_q$ changes abruptly in just a few steps in the case of $n_0=12$, indicating the existence of cascading of majority-rule process.}\label{fig2}
\end{figure}

In Fig.~\ref{fig1}, we plot the average exit probability for systems with $G=9$ and critical masses $n_0=5$, $6$ and $8$. We see that with increasing $n_0$, the system can, up to a certain time, more easily reach consensus. This phenomenon can be understood as follows. When the critical mass $n_0$ is small, it is likely that both opinions are in local majority in different regions of the lattice. The minority opinion within these regions vanishes so that there will be contiguous regions of one opinion. Since the system, in such a situation, only can evolve in the interface between the opinions the relaxation toward consensus will slow down. A larger critical mass means that more information is integrated before a decision is taken, so that the probability that the global minority will form a consensus region decreases, and thus the time to reach consensus decreases.

We increase the local interaction scale to $G=13$ in Fig.~\ref{fig2}. We use critical masses $n_0=7$, $11$, and $12$ to illustrate the behavior. As for the smaller local scale, we find that the capability of reaching consensus increases with the critical mass. However, for $n_0=12$, $E_q$ exhibits a conspicuously different behavior. In this case, $E_q$ close nearly zero for $t<20$ and approaches unity quickly in the interval $25\lesssim t\lesssim 30$. To sketch the dynamics picture behind this phenomenon we note that, for large critical masses the system first spends most time reshuffling the opinion locally while the majority-rule is almost never applied. Then, when the majority-rule is first applied, which is likely to happen around $t=25$, the system converges to consensus in just a few steps (which is illustrated explicitly below).

\begin{figure}
\includegraphics[width=0.9\linewidth]{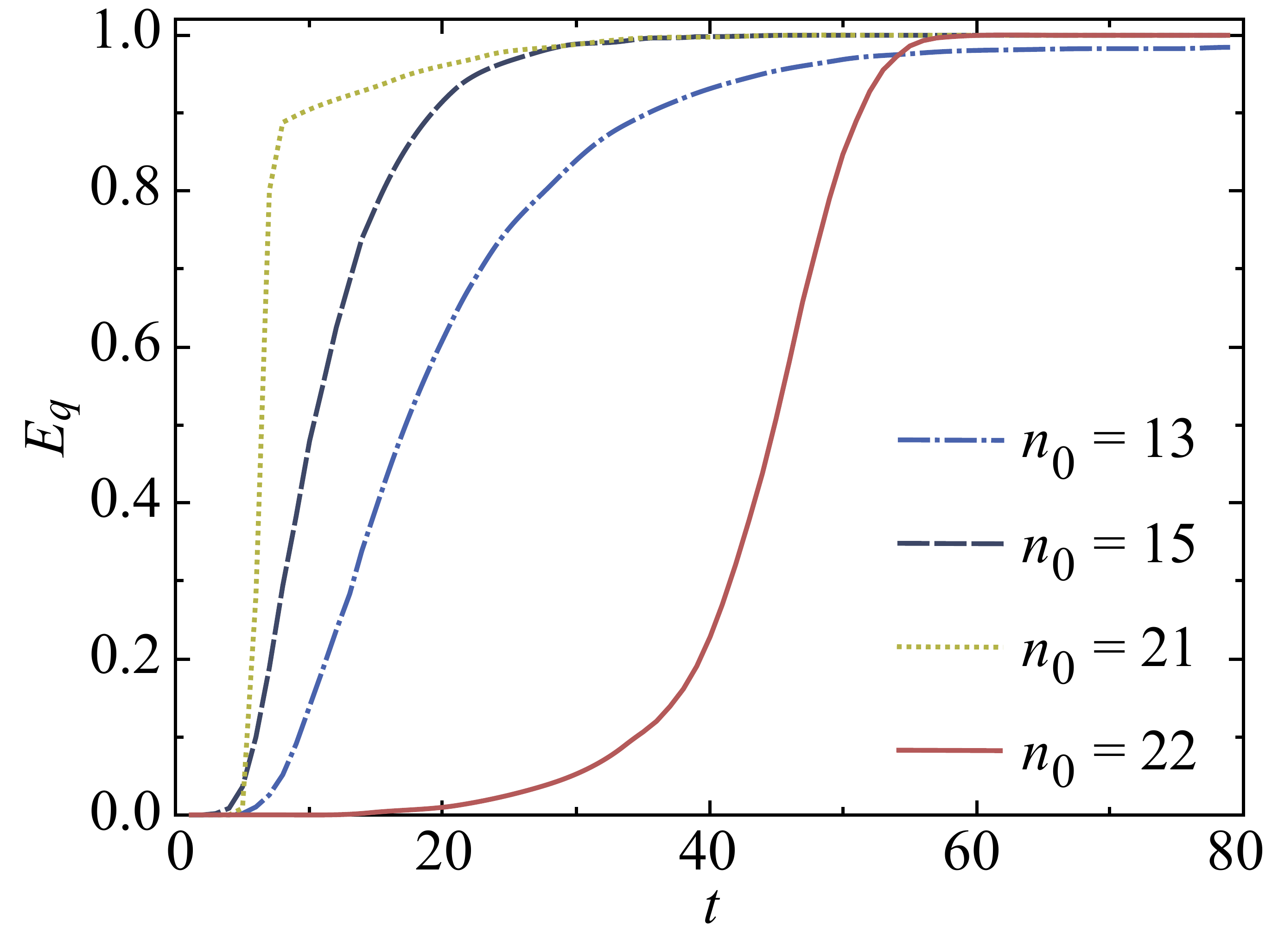}
\caption{(Color) Average exit probability as a function of the relaxation time, similar to Fig.~\ref{fig1}, but for a local-influence group of size $G=25$ and critical mass $n_0=13$, $15$, $21$, and $22$.}\label{fig3}
\end{figure}

\begin{figure}
\includegraphics[width=0.9\linewidth]{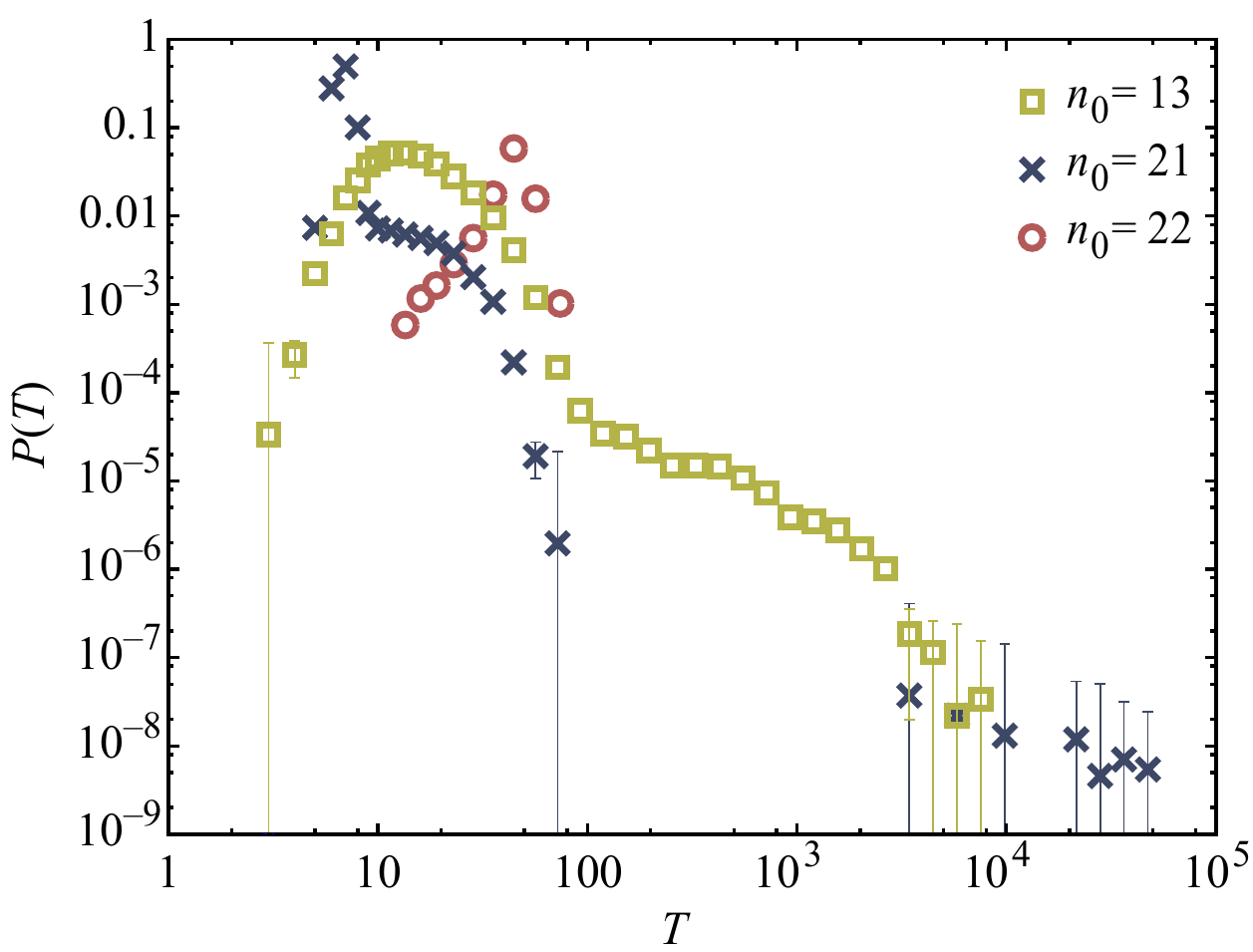}
\caption{(Color) The probability mass function of consensus times for the system with $G=25$ and critical masses $n_0=13$, $21$, and $22$. The data is plotted log-binned in log-log scale to high-light the tail behavior. We note that the long tail for $n_0=13$ displays approximately exponential decay, similar to those have been noted before in Ref.~\cite{Chen2005pre} where a simple majority rule model is considered. Error bars represent the standard error (and omitted if smaller than the symbol size).}\label{fig4}
\end{figure}

To check the universality of such a dynamic phenomenon, we continue to an even larger interaction-group sizes, see Fig.~\ref{fig3}. Also in this case we see that the propensity of reaching consensus increases with increasing $n_0$, and that this increase saturates an intermediate value (but closer to one than zero). With even growth of the critical mass, the evolution of $E_q$ behaves similar to the $n_0=12$-case of Fig.~\ref{fig2}. In other words, for a transient, the system rearranges the opinion configurations, and after which the majority opinion quickly overtakes the entire system (this will be illustrated below). In light of these results, we argue that there exists optimal value of the critical mass for the system to reach consensus, which depends closely on the size of the interaction group. For $G$ less than a number between $9$ and $12$, the optimal mass is $G-1$, while for large $G$ the optimal critical mass for consensus formation occur at some intermediate value of $n_0$ between $G/2$ and $G$.

\begin{figure}
\includegraphics[width=0.9\linewidth]{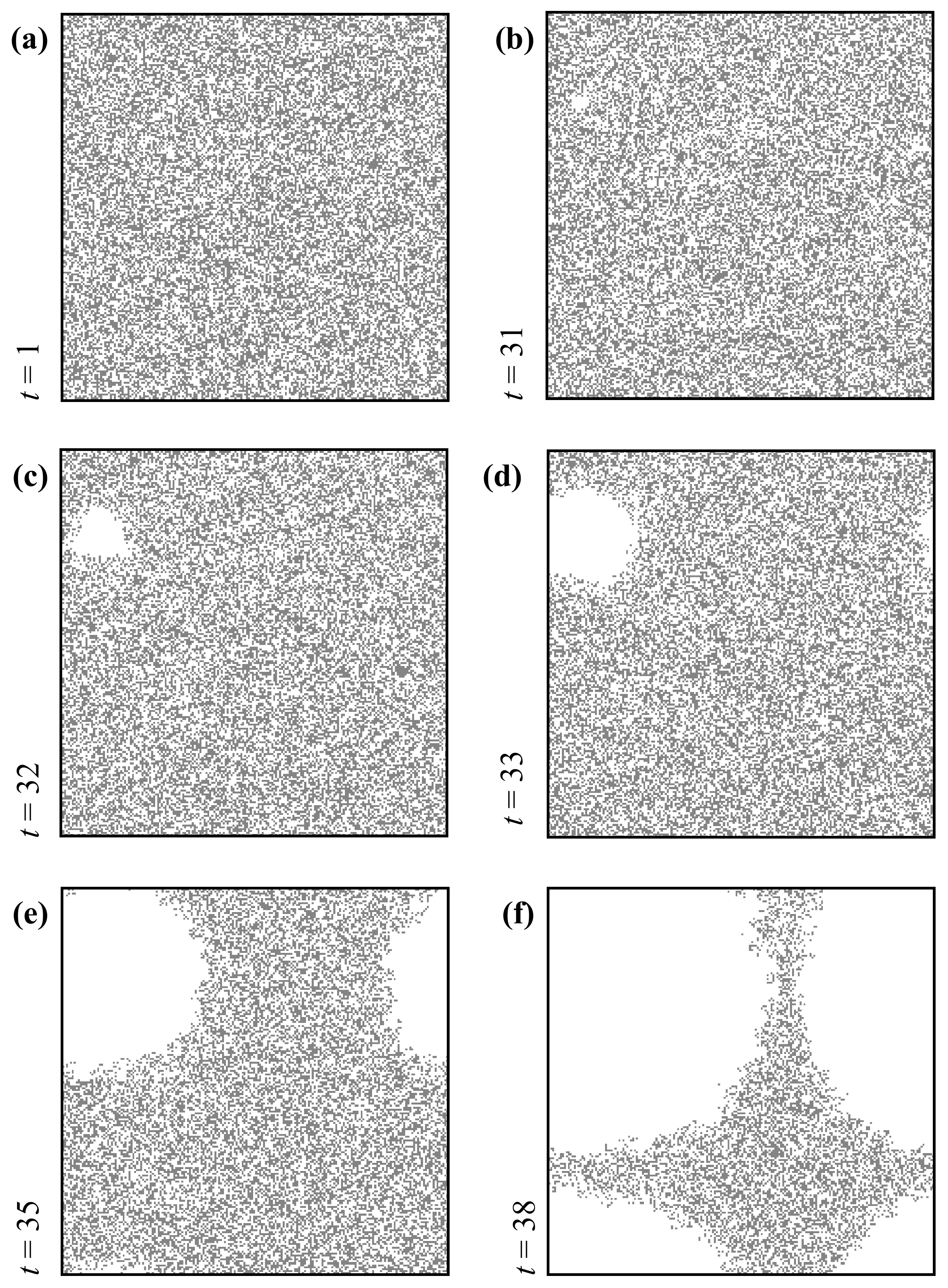}
\caption{Snapshots of opinion configuration of the agents at different relaxation time $t$ in a typical realization of our threshold majority rule dynamics. The white and black sites correspond, respectively, to the agents holding opinion $+1$ or $-1$. The size of local-influence group $G=25$, and the critical mass $n_0=22$.}\label{fig5}
\end{figure}

To further explore the dynamic properties of our model, in Fig.~\ref{fig4} we plot the probability mass function of consensus times $p(T)$ for $G=25$ and three different $n_0$s. It is shown that for small values of the critical mass, $p(T)$ displays similar properties as were found in~\cite{Chen2005pre}, where $p(T)$ involves two separate time scales. We see this in the coexistence of peaks and long tails in the distribution. The peak value of the distribution corresponding to the most probable consensus time~\cite{Chen2005pre,Li2006pre}, while the long tail arises from situations where which domains of opposite opinions organize into metastable configurations. As stated before, it is possible for the minority to form domains at local scale given that $G$ and $n_0$ are not too large. Once such domains appear, the system need extremely long time to reach global consensus. Peculiarly, in the case of sufficiently large critical mass (here $n_0=22$), we find that the longer time scale vanishes (see Fig.~\ref{fig4}). We therefore posit that under such conditions, the global minority opinion cannot evolve to form big groups, hence the metastable states cannot be formed either. In such situation, however, the system needs much time to rearrange the configuration a state instable to the majority-rule process. We have verified that this scenario holds qualitatively for even larger values of $n_0$.

Finally, to get an intuitive impression of the process of opinion formation in the case of large critical masses, in Fig.~\ref{fig5} we plot the actual opinion configurations of a run with $G=25$ and $n_0=22$. The configuration does not change much until $t=31$, when a small group of agents reach local consensus in the upper-left part of the grid. After this nucleation of opinions, the cascading of majority-rule process starts, leading to consensus after just a few time steps. These snapshots corroborate our previous picture of the dynamics for large $n_0$. The snapshots for $n_0=22$ are quite different with $n_0=13$ where different opinions form local domains that competes  with each other for some time (results not shown here). This situation resembles the study in Ref.~\cite{Chen2005pre}.

\section{Conclusions}\label{conclusion}

In conclusion, we have studied a threshold majority-rule model of opinion formation, wherein the agents can take one of two opinions and the evolution of opinion is implemented by group updating. In particular, all agents inside a group will follow the majority provided that the number in majority exceeds a threshold value. This is to model winner-takes-it-all phenomena in social systems. If the majority is too small, agents just exchange their opinion randomly. By studying the exit probability of the system as a function of the relaxation time, we find that there is an optimal critical mass in terms of the average time it takes for the system to reach consensus. The optimal critical mass depends strongly on the size of interaction group. In particular, in our threshold majority-rule model, the longer time scale found in the distribution of consensus time for small critical masses disappears with more strict conditions for achieving consensus is assumed.

The obvious step towards increased realism is to consider opinion spreading on contact structures more like empirical social networks. Due to the varying size of the neighborhoods in real-world networks it is not straightforward to control the size of the local influence group. In general we hope this work inspire deeper investigations into systems with varying scales of the local interaction.

\acknowledgments{
This work is supported by the Swedish Research
Council (Z.X.W. and P.H.), the Swedish Foundation for Strategic
Research (P.H.), and the WCU program through NRF Korea funded by MEST (R31--2008--000--10029--0).}

\bibliographystyle{h-physrev3}

\end{document}